# Design of reconfigurable Huygens metasurfaces based on Drude-like scatterers operating in the epsilon-negative regime

Alessio Monti[1,*], Stefano Vellucci[2,3], Mirko Barbuto[2,3], Luca Stefanini[1], Davide Ramaccia[1], Alessandro Toscano[1], and Filiberto Bilotti[1,3]

[1] *ROMA TRE University, Department of Industrial, Electronic and Mechanical Engineering, Via Vito Volterra 62, 00146, Rome, Italy.*

[2] *Niccolò Cusano University, Department of Engineering, Via don Carlo Gnocchi 3, 00166, Rome, Italy*

[3] *Virtual Institute for Artificial Electromagnetic Materials and Metamaterials, Place du Levant, 3, 1348 Louvain-la-Neuve, Belgium*

\* *Corresponding author: Alessio Monti, alessio.monti@uniroma3.it*

***Abstract*** **– In this study, we investigate the feasibility of designing reconfigurable transmitting metasurfaces through the use of Drude-like scatterers with purely electric response. Theoretical and numerical analyses are provided to demonstrate that the response of spherical Drude-like scatterers can be tailored to achieve complete transmission, satisfying a generalized Kerker's condition at half of their plasma frequency. This phenomenon, which arises from the co-excitation of the electric dipole and the electric quadrupole within the scatterer, also exhibits moderate broadband performance. Subsequently, we present the application of these particles as meta-atoms in the design of reconfigurable multipolar Huygens metasurfaces, outlining the technical prerequisites for achieving effective beam-steering capabilities. Finally, we explore a plausible implementation of these low-loss Drude-like scatterers at microwave frequencies using plasma discharges. Our findings propose an alternative avenue for Huygens metasurface designs, distinct from established approaches relying on dipolar meta-atoms or on core-shell geometries. Unlike these conventional methods, our approach fosters seamless integration of reconfigurability strategies in beam-steering devices.**

## I. INTRODUCTION

Metasurfaces, consisting of a periodic arrangement of sub-wavelength and electrically thin scatterers engineered to provide unusual and useful electromagnetic response, have emerged as a revolutionary platform in the realm of electromagnetic wave manipulation, offering unprecedented control over propagation and scattering [1]. Depending on the complexity of the fundamental scatterers (also referred to as *meta-atoms*) and, specifically, on the nature of the currents induced on them by an incident plane wave, metasurfaces may enable a wide plethora of unprecedented physical effects, such as advanced polarization, scattering, and absorption control [2]-[5], invisibility and camouflage [6],[7], beam-steering and beam-forming [8]-[11], wavefront manipulation [13], etc.

Among the various trends in this field, reconfigurability [14]-[15] has become a focal point of the research efforts, as it promises dynamic and programmable control of the metasurface macroscopic behavior, depending on the evolving requirements of the system. For instance, it is expected that reconfigurable metasurfaces will play a fundamental role in future wireless systems given their ability to enable dynamic control over the direction of transmitted or received beams and to transform the environment into an active element of the communication system [16]-[17]. Similarly, optically-reconfigurable metasurfaces may increase the resolution of imaging systems [18], allow dynamic control over the focus and depth of field in microscopy [19], improve spectroscopy techniques [20], play a relevant role in manipulating quantum states of light in quantum communication and quantum computing applications [21], etc.

Among the different families of metasurfaces, the so-called Huygens (or refractive) metasurfaces [22]-[23] play a relevant role as they are widely used to implement beam-steering devices working in transmission mode. The meta-atoms of these structures are engineered in such a way that they provide full-transmission (*e.g*., magnitude of the transmission coefficient close to one at the frequency of operation) and a desired phase-shift through the metasurface layer. To achieve this effect, different approaches are possible: the most consolidated one relies on the use of meta-atoms that, when excited by an external field, sustain both electric and magnetic currents. By acting on the geometry of the meta-atom, the electric and magnetic response can be balanced at a desired frequency to achieve full-transmission and a desired phase shift. This approach is widely used at microwave frequencies, for instance through the use of a layout made by cascaded reactive sheets [9]-[12],[22], or at optical frequencies, through the use of Mie resonators, *i.e.,* dielectric scatterers with high refractive index, whose geometry is engineered to achieve a proper balance between the amplitude and phase of the electric dipole (ED) and magnetic dipole (MD) [25]-[29].

More recently, however, it has been shown that Huygens behavior can be also achieved considering a proper combination of multipoles excited within a small particle [30],[31]. Scatterers designed to exhibit this effect, also referred to as generalized Kerker's effect, can be used as meta-atoms for designing *multipolar Huygens metasurfaces*, showing several advantages over dipolar ones [32],[33]. For instance, in [34], it has been shown how dipolar-quadrupolar Huygens metasurfaces can ensure a wider phase coverage when sweeping only the lattice constant. In [35], it has been discussed how to achieve a high-Q passband filter relying on the proper combination of dipoles and quadrupoles. In both these works, the multipolar metasurface is realized through core-shell geometries involving both high-index and low-index dielectrics. Such a configuration, therefore, cannot easily accommodate for reconfigurability strategies, limiting the possible application of multipolar metasurfaces to the static scenarios.

In this contribution, we explore a different approach for the design of multipolar Huygens metasurfaces based on reconfigurable Drude-like scatterers. Through a combination of theoretical and numerical tools, we first derive and explain the conditions under which such scatterers can be used as meta-atoms for Huygens metasurfaces; then, we show how the interaction occurring in the periodic environment is able ensuring the full phase coverage of the transmitting metasurface. Finally, we assess the reconfigurability performance and tolerance to material losses of such devices and discuss a possible implementation with realistic materials in the microwave regime.

## II. ANALYSIS AND DESIGN OF THE INDIVIDUAL SCATTERER

The fundamental structure considered in this work is shown in Fig. 1(a) and consists of a sphere made of Drude-like material. We assume that the sphere is in free-space and illuminated by an external normally-incidence plane wave. Given the symmetry of the particle, the discussion below applies to any possible incident polarization. The radius of the sphere is denoted with $a$, while its permittivity is modelled through the following frequency-dependent function (Drude model):

$$\varepsilon_s = 1 - \frac{\omega_p^2}{\omega^2 - j\omega\gamma}, \qquad (1)$$

being $\omega_p$ and $\gamma$ the plasma and collision frequency, respectively. In this Section, we consider the almost-lossless scenario and fix $\gamma = 10^{-3}\omega_p$. We will discuss later in the manuscript the effects of realistic material losses on the performance of devices based on these structures.

The total scattering cross section of the sphere can be expressed in summation form by using the well-known Mie theory [36]:

$$Q_{sca} = \frac{2}{x^2}\sum_{n=1}^{\infty}(2n+1)(|a_n|^2 + |b_n|^2) \qquad (2)$$

where $a_n$ and $b_n$ are the scattering coefficients, which read as:

$$a_n = \frac{m\psi_n(mx)\psi'_n(x) - \psi_n(x)\psi'_n(mx)}{m\psi_n(mx)\xi'_n(x) - \psi_n(x)\xi'_n(mx)}, \qquad (3)$$

$$b_n = \frac{m\psi_n(mx)\psi'_n(x) - \psi_n(x)\psi'_n(mx)}{m\psi_n(mx)\xi'_n(x) - \psi_n(x)\xi'_n(mx)}. \qquad (4)$$

In the above expressions, $\psi_n$ and $\xi_n$ denote the Riccati-Bessel functions, $m = \sqrt{\varepsilon_s}$ is the refractive index of the sphere and $x = ka$, being $k$ the free-space wavenumber. Other useful parameters to describe the scattering by the sphere are the so-called forward and backward scattering cross sections, which quantify the scattering amplitude in the forward ($z > 0$) and backward ($z < 0$) direction, respectively. These quantities can be expressed as follows:

$$Q_{fs} = \frac{1}{x^2}\sum_{n=1}^{\infty}(2n+1)(a_n + b_n) \qquad (5)$$

$$Q_{bs} = \frac{1}{x^2}\sum_{n=1}^{\infty}(2n+1)(-1)^n(a_n - b_n) \quad (6)$$

Consequently, the front-to-back (F/B) ratio can be expressed as:

$$F/B = \frac{Q_{fs}}{Q_{bs}} \quad (6)$$

A potential candidate scatterer for implementing Huygens metasurfaces needs to satisfy two fundamental requirements: *i)* high F/B ratio (*i.e.,* this is also referred to as Huygens behavior), possibly within a large bandwidth, *ii)* high total scattering cross section (otherwise, the low backscattering would be a direct consequence of the absence of the interaction with the impinging field, which is a trivial case for low backscattering) [37],[38].

We first focus on the second requirement and look at the design conditions for which a generic Drude scatterer does have a strong interaction with the impinging field. In Fig. 1(b), we report the total scattering cross section, expressed by eq. (2), of the spherical particles for different frequencies and for different size. In particular, the *x*-axis of the contour plot represents the size parameter $F$, defined as $a = \lambda_p/F$, while the *y*-axis represents the frequency parameter *x*, defined as $f = xf_p$, being $f_p = \omega_p/(2\pi)$. In other terms, $F$ refers to the electrical length of the radius of the particles, while *x* is the operation frequency normalized to the plasma frequency of the scatterer material. The scale of the contour plot has been chosen to show only values of $Q_{sca}$ higher than 8 *dB,* which results in a strong interaction of the scatterer with the impinging electromagnetic field.

As it can be appreciated, in the frequency region $0 < x < 1$, there are several size combinations for which $Q_{sca} > 8dB$. We denote with Region#1 the one for which $10 < F < 8$, Region#2 the one for which $4 < F < 6$ and, finally, Region#3 the one for which $4 < F < 6$. Conversely, for $x > 1$, there is no size region of the plot for which the particle reaches a satisfactory level of the total scattering. This behavior can be easily understood by considering eq. (1) and, in particular, the behavior of the permittivity function beyond the plasma frequency (*i.e.*, for $x > 1$). In this frequency range, the permittivity of the particle is in the range (0,1) and asymptotically approaches the one of the free-space, resulting in almost negligible wave-matter interaction. Therefore, we can conclude that the only frequency region for which a Drude scatterer can be suited for implementing Huygens metasurfaces is the epsilon-negative (ENG) one.

According to the outcome of the total scattering analysis, we have investigated the F/B ratio of the spherical particles in the three identified Regions to assess which one is more promising for implementing Huygens metasurfaces. The results are shown in Fig. 2. As it can be appreciated, the scatterers belonging to Region#1 show a F/B ratio close to unitary (0 *dB*), meaning that the forward and backward scattering assume almost the same values. Therefore, such scatterers are not a viable option for implementing Huygens metasurfaces. Conversely, the scatterers belonging to Region#2 and Region#3 exhibit a strong F/B ratio around $x \approx 0.61$ and $x \approx 0.50$, respectively. The stronger F/B ratio is achieved in Region#2, for $F \approx 6.0$; however, its bandwidth is quite narrow. Conversely, the scatterers in Region#3, and in particular for $F \approx 2.0$, keeps their F/B ratio higher than 6 $dB$ within a quite broad range of frequencies, *i.e.*, for $0.43 < x < 0.53$.

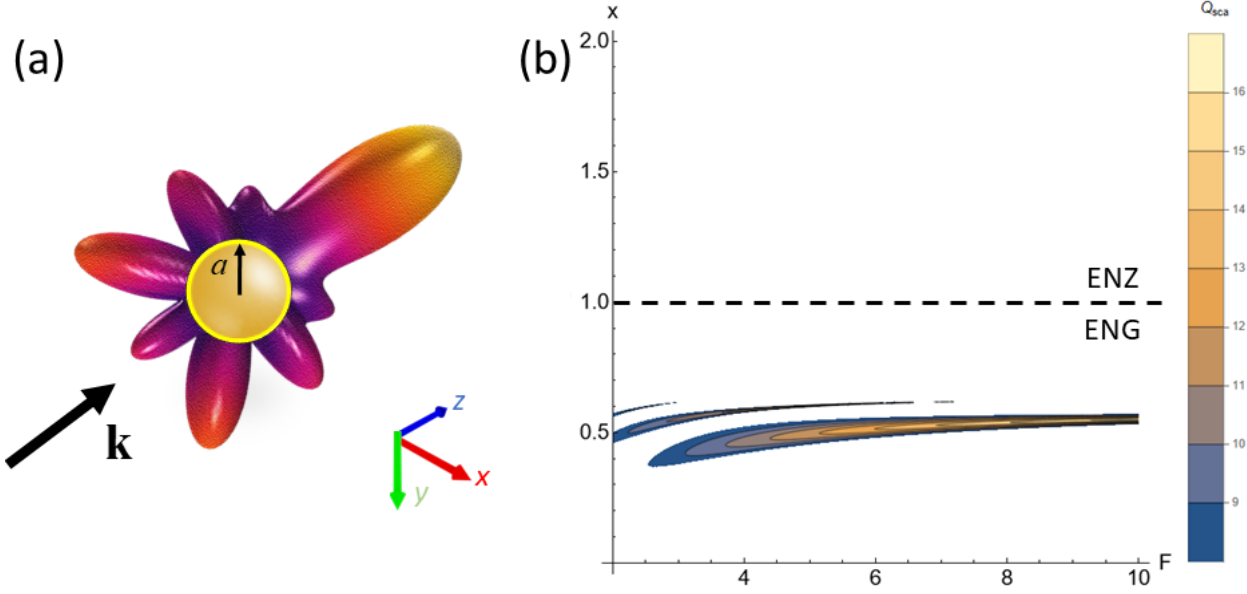

Fig. 1. (a) Drude-like spherical scatterer illuminated by a normally impinging plane wave. The superimposed scattering pattern is arbitrary and chosen only for illustrative purposes. (b) Scattering cross section $Q_{sca}$ (expressed in $dB$) of the Drude-like scatterer *vs.* the size parameter $F$ and the frequency parameter $x$. The scale of the plot has been chosen to show only the values of $Q_{sca}$ higher than 8 $dB$. The horizontal dashed line separates the frequency region where the permittivity of the Drude scatterer is epsilon-near-zero (ENZ) or epsilon-negative (ENG).

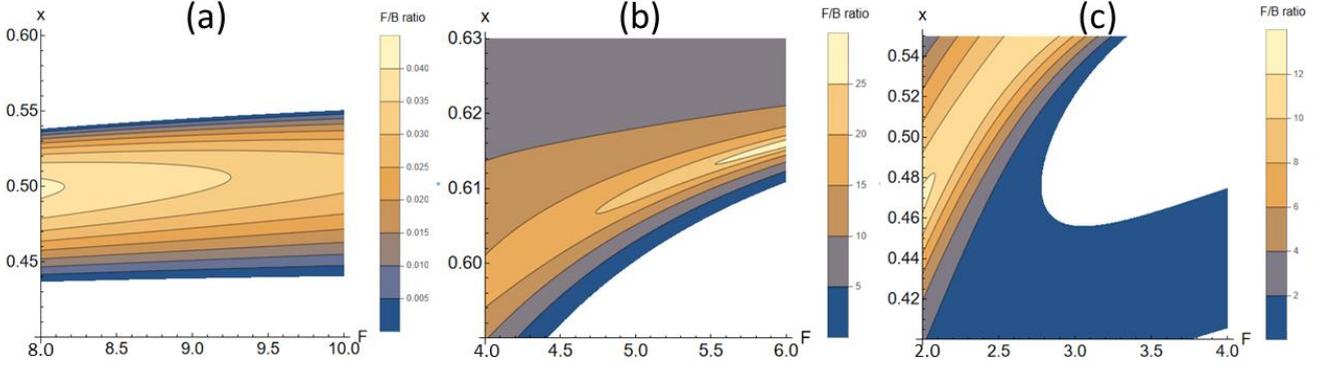

Fig. 2. F/B ratio (expressed in $dB$) of the Drude-like spherical scatterer in the different identified size regions.

It could be, therefore, concluded that the best size parameter for which a Drude-like scatterer can be used for implementing Huygens metasurfaces is $F \approx 2.0$, corresponding to a radius $a = \lambda_p/2$.

To get additional physical insights, we have explored the frequency behavior of the scattering coefficients of a Drude-like spherical scatterer with $a = \lambda_p/2$ in its ENG frequency region, *i.e.*, $0 < x < 1$. According to the well-known rule-of-thumb about the significance of the Mie scattering coefficients, we limit our analysis to the multipole $N_{max} = [ka] = 4$. The amplitude of the scattering coefficients is shown in Fig. 3(a), while in Fig. 3(b) we show the total, forward and backward scattering cross section in the same frequency range. As it can be appreciated, around $x = 0.50$, the magnitudes of the electric dipole and quadrupole hit their maximum values and keep approximately constant in a moderately wide frequency range (within which the other scattering parameters have negligible amplitude). As discussed in [31], this scenario corresponds to one of the so-called generalized Kerker's conditions, for which the scatterer exhibits almost full forward-scattering. This is confirmed by the results shown in Fig. 3(b): at around half of the plasma frequency ($x = 0.50$), there is a wide peak of $Q_{fs}$ and a negative peak of $Q_{bs}$. In this frequency region, the magnitude of $Q_{sca}$ keeps above $6\ dB$, confirming a strong interaction of the impinging wave with the scatterer. Another frequency range where similar conditions apply, caused by the interaction between the electric quadrupole and the electric octupole, can be observed at a higher frequency, slightly before $x = 0.60$. In this case, however, the bandwidth is significantly narrower and this may prevent achieving a full phase coverage when moving from the single scatterer to the periodic metasurface. We underline that in Fig. 3(b) we also show the results of numerical simulations carried out through a FEM-based simulation algorithm, that confirm an excellent agreement between theoretical and numerical values.

From the above analysis, we can conclude that Drude-like scatterers with a radius equal to half of their plasma wavelength do exhibit a moderately broadband Huygens behavior. This effect is the result of the proper co-excitation of electric and magnetic quadrupoles and owns all the features required for Huygens metasurface design.

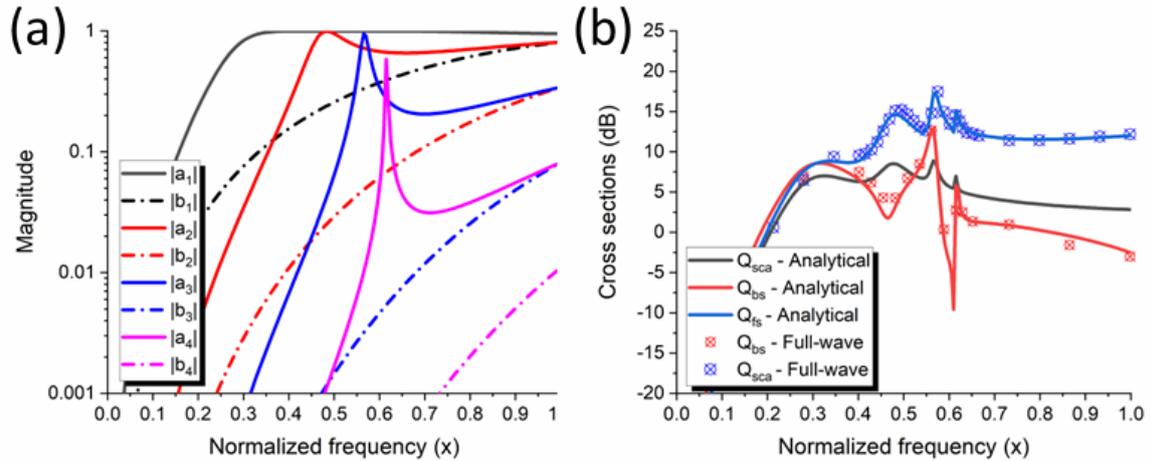

Fig. 3. (a) Magnitude of the first four scattering coefficients *vs.* the frequency parameter *x* for a Drude-like scatterer with $a = \lambda_p/2$. (b) Total, backward, and forward scattering cross section of the same particle.

III. ANALYSIS OF THE HOMOGENOUS METASURFACE

We consider now the electromagnetic behavior of a homogenous metasurface, shown in Fig. 4(a), consisting of a periodic arrangement of Drude-like scatterers. According to the analysis carried out on the individual scatterer, the size of the meta-atom is set to $a = \lambda_p/2$ and the transmission coefficient is investigated in the frequency range around the generalized Kerker's condition, *i.e.*, $x = 0.5$. It is worth noticing that, at the operative frequency, the size of the meta-atom is $a = \lambda_0/4$, which is not deeply subwavelength but still reasonably small compared to the operation wavelength;

further miniaturization (if needed) can be achieved by using a dielectric shell with permittivity substantial higher than one encapsulating the Drude-like scatterer.

The analysis has been carried out through numerical simulations and for different values of the inter-element separation distance. For numerical simulations, we have set an arbitrary value of the plasma frequency ($f_p = 7.51\ GHz$). Consequently, the radius of the Drude-like scatterer is approximately equal to 20 *mm*. As it can be appreciated in Fig. 4(b), there is a wide range of frequencies for which the magnitude of the transmission coefficient of the metasurface keeps above $0.7\ (-3\ dB)$. This range spans around the frequency at which the individual scatterer hits the Kerker's condition, and its bandwidth depends on the inter-element separation distance. This effect is a consequence of the different interaction between adjacent particles and has been earlier discussed for dipolar metasurfaces [28]. In this case, the value of the separation distance that optimizes the quality factors of the two resonances and returns the larger transmission bandwidth is $d = 3.00a$, corresponding to the normalized frequency bandwidth $0.36 < x < 0.55$; within this range, the phase of the transmission coefficient almost covers the entire $2\pi$.

To check the possibility to use such a structure as a Huygens metasurface, we have investigated its reconfigurability performance. In particular, we have fixed the operation frequency $f_0$ at $x = 0.5$ (corresponding to $f_0 = 3.8\ GHz$ in this example) and sweep the plasma frequency $f_p$ of the Drude-like scatterers. The magnitude and phase of the transmission coefficient *vs.* the plasma frequency are reported in Fig. 5. In these plots, the *x*-axis is $\tilde{f}_p$ defined as the plasma frequency normalized to its original value. As it can be appreciated, the magnitude of the transmission coefficient keeps above $0.7\ (-3\ dB)$ for $0.2 < \tilde{f}_p < 0.75$ and for $\tilde{f}_p > 0.86$. Within these ranges, an almost full $2\pi$ phase coverage is achieved, *i.e.*, $-180° < arg(T) < -100°$ and $-127° < arg(T) < 140°$, respectively.

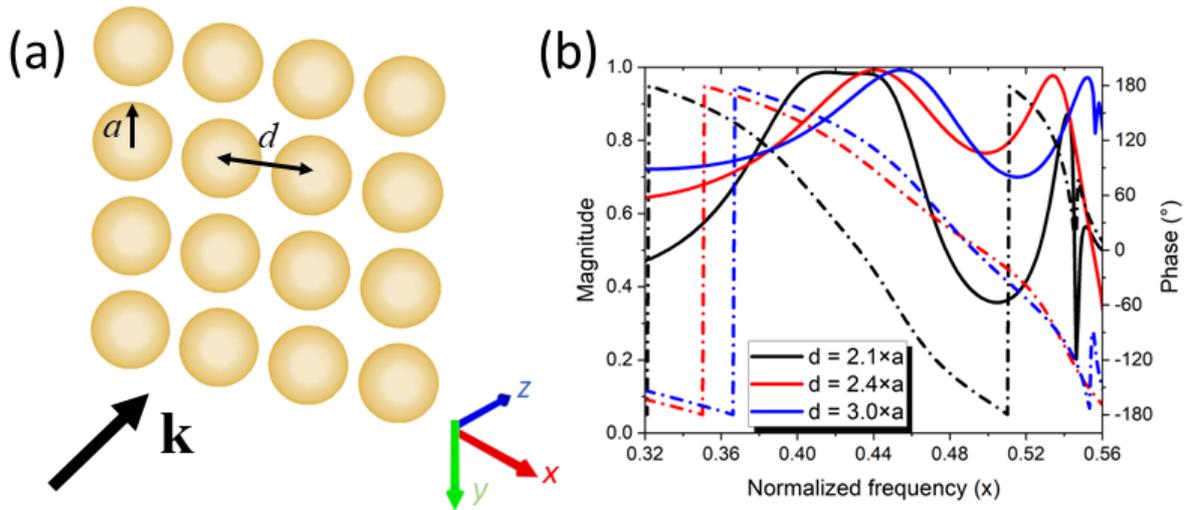

Fig. 4. (a) Geometry of the homogenous metasurface. (b) Transmission coefficient of the homogenous metasurfaces for different inter-element separation distance. Continuous and dashed lines refer to magnitude and phase of the transmission coefficient, respectively.

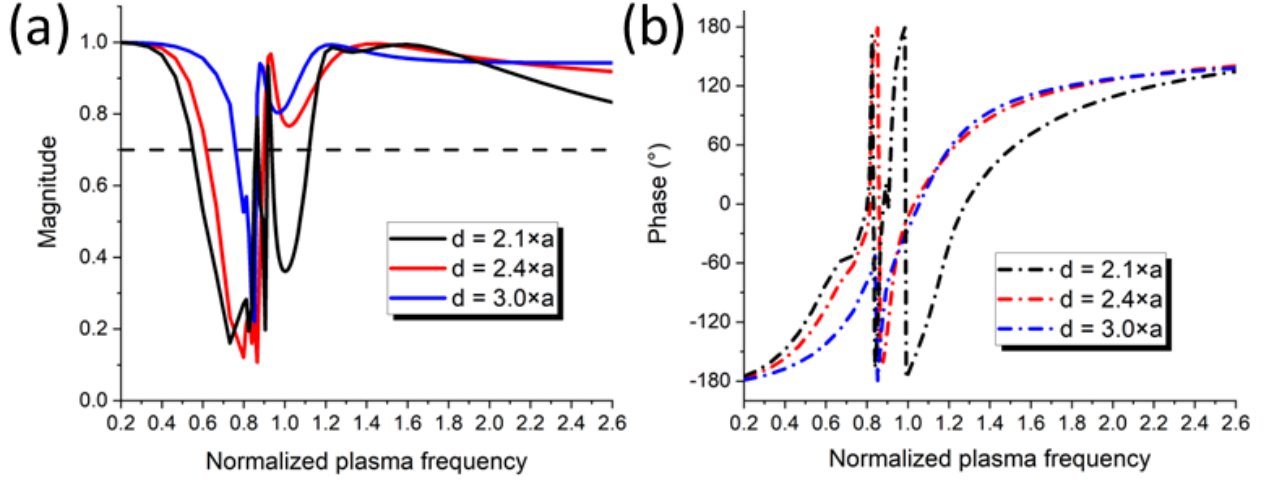

Fig. 5. (a) Magnitude and (b) phase of the transmission coefficient at $x = 0.5$ for different values of the normalized plasma frequency $\widetilde{f_p}$.

We can therefore conclude that Drude-like scatterers, designed to satisfy the generalized Kerker's condition at half of their plasma frequency, can be effectively used as fundamental meta-atom for reconfigurable refractive Huygens metasurfaces. The full phase coverage is achieved when the plasma frequency is swept between 20% and 260 % of the original value used to size the particle. In the next Section, we will describe how this result can be exploited to design a reconfigurable refractive metasurface based on Drude-like scatterers.

IV. DESIGN AND NUMERICAL RESULTS OF THE RECONFIGURABLE GRADIENT METASURFACE

We consider here the design of a phase-gradient metasurface consisting of four Drude-like scatterers with different plasma frequencies $f_{p1}$, $f_{p2}$, $f_{p3}$, and $f_{p4}$ illuminated by a normally-incident plane-wave (inset of Fig. 6(a)). As initial value of the plasma frequency, we set $f_p = 7.51\ GHz$ (as in the previous Section). Hence, the radius of the Drude-like scatterer is approximately equal to 20 *mm*. The inter-element separation distance is set to $d = 2.24a = 45\ mm$. The period of the gradient metasurface is equal to $p_{MTS} = 4d = 180\ mm$.

We consider two possible states of the reconfigurable metasurface. The first state (*State 1*) is the transparent one, in which all the plasma frequencies of the four scatterers are set to the same values, *i.e.*, $f_{p1} = f_{p2} = f_{p3} = f_{p4} = f_{tran}$. In this scenario, there is no phase variation along the *x* direction and, therefore, we expect full transmission by the metasurface in the normal direction. In the second state (*State 2*), instead, each scatterer is reconfigured to have a different plasma frequency and, in particular, to introduce a gradient of the transmission phase along the *x*-direction. The optimal normalized plasma frequency values, originally guessed through a local-approach and, then, corrected through a full-wave optimization are, $\widetilde{f_{p1}} = 0.52$, $\widetilde{f_{p2}} = 1.68$, $\widetilde{f_{p3}} = 1.34$, and $\widetilde{f_{p4}} = 1.02$. According to the well-known grating formula, the expected refraction angle by the metasurface in the second state is:

$$\vartheta_t = \operatorname{asin}\left[\frac{\lambda_0 + p_{MTS}\sin(\theta_{in})}{p_{MTS}}\right] = \operatorname{asin}\left(\frac{\lambda_0}{p_{MTS}}\right) = 26° \quad (7)$$

In Fig. 6(a), we report the magnitude of the most relevant Floquet-Bloch (FB) transmission coefficients in State 1 and State 2. As it can be appreciated, in State 1, the magnitude of the $S_{11}$ parameter is almost equal to $0\ dB$ within a wide range of frequencies, showing that the uniform metasurface is almost transparent to the impinging field. Fig. 6(b), which reports the electric field profile in State 1, confirms this conclusion. In State 2, instead, most of the impinging energy is coupled to the FB mode aligned along the transmission angle $\vartheta_t$ expressed by eq. (7). Indeed, in the electric field profile shown in Fig. 6(c), it is possible to appreciate the high-efficiency anomalous refraction introduced by the gradient metasurface.

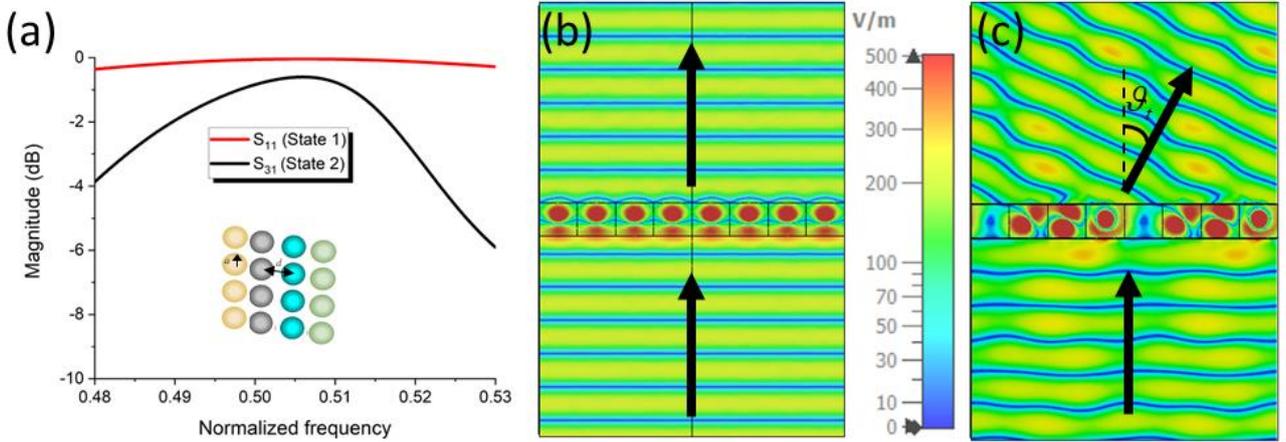

Fig. 6. (a) Magnitude of the transmission coefficient for different Floquet modes in State 1 and State 2. In the inset, it is shown a picture of the gradient metasurface. (b)-(c) Electric field maps in State 1 and State 2, respectively.

Additional metasurface states and, therefore, multiple steering angles can be achieved by using a higher number of independent reconfigurable meta-atoms; however, this goes beyond the scope of this paper. Nonetheless, the simple example discussed above confirms that properly designed Drude-like scatterers with tunable plasma frequency are an effective route for designing reconfigurable refractive metasurfaces.

Before concluding this Section, we discuss the sensitivity of this design to electromagnetic losses and variation of the angle of incidence. For the former aspect, we have evaluated the magnitude of $S_{31}$ in State 2 for different values of the collision frequency $\gamma$. In particular, the collision frequency of each scatterer has been expressed as a function of the respective plasma frequency, i.e., $\gamma = \gamma_x \omega_p$. The results for different values of $\gamma_x$ are shown in Fig. 7(a). As it can be appreciated, material losses cause both a frequency shift of the peak of the transmission coefficient and a reduction of its maximum value. Assuming a $0.7\ (-3\ dB)$ threshold, we can conclude that the maximum level of losses that can be tolerated is $\gamma_x = 0.03$, i.e., the collision frequency should not exceed 3% of the plasma frequency. This value can be considered as conservative, since the values of the plasma frequency of each meta-atom have not been optimized again for each level of losses (the original values, obtained in the almost lossless scenario, have been used for each simulation).

Regarding the behavior with the angle of incidence, in Fig. 7(b) we report the magnitude of $S_{31}$ for different values of $\vartheta_{in}$. As it can be appreciated, the performance keeps satisfactorily up to 40° because of the angular stability of the meta-atom. For higher values, performance deteriorates quickly because the period of the gradient metasurface is such that eq. (7) is not satisfied anymore. This should not be considered as a limitation of the proposed meta-atom but, rather, as a fundamental limitation of gradient-metasurfaces [39].

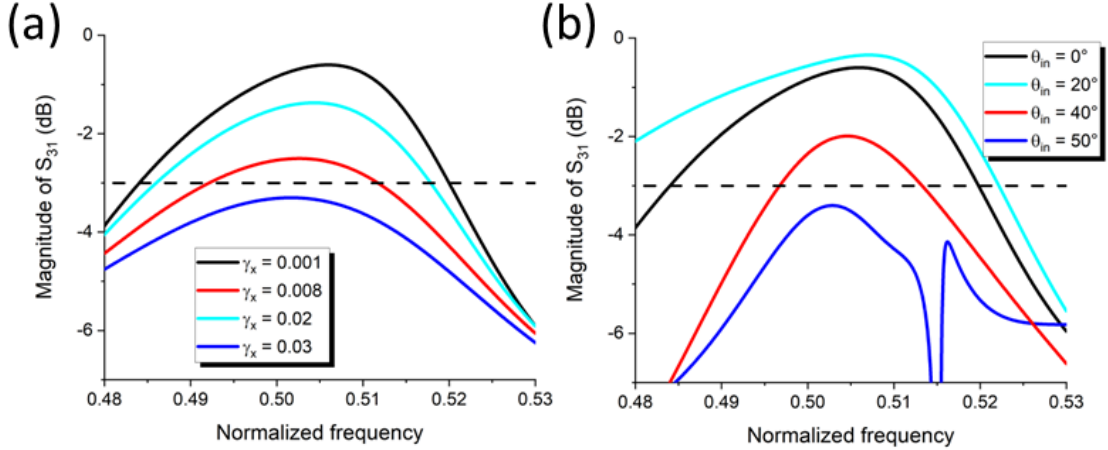

Fig. 7. Performance of the gradient metasurface (State 2) vs. (a) material losses and (b) angle of incidence. The horizontal dashed line represents the -3 *dB* threshold.

## V. DISCUSSION AND CONCLUSIONS

In this work, we have described a possible route for designing reconfigurable refractive metasurfaces based on Drude-like scatterers. Specifically, we have shown how such scatterers can be designed to achieve a broadband generalized Kerker's condition due to the proper superposition of electric multipoles. It has also been shown that, by changing the plasma frequency of the scatterers in a periodic configuration, it is possible to achieve an almost complete phase coverage with high transmission efficiency; this effect has been finally exploited to design a simple reconfigurable refractive metasurface.

The proposed technique can be considered as alternative to other well-established approaches based on the use of dipolar meta-atoms, which require the excitation of a magnetic response inside the meta-atom. The proposed approach only requires meta-atom with an electric response and, therefore, may enable the use of simpler reconfiguration mechanisms.

To summarize, the approach discussed in this paper is based on the use of spherical Drude-like scatterers with reconfigurable plasma frequency. Once set the original plasma frequency to an arbitrary value $f_p$, the radius of the required particle should be $a = \lambda_p/2$ and the Huygens behavior is obtained for an operation frequency around $f_0 = f_p/2$. The almost complete phase coverage can be obtained by changing the plasma frequency in the range $0.2f_p - 2.6f_p$. The transmission efficiency keeps above the $-3dB$ threshold as far as the collision frequency does not exceed 3% of the plasma frequency of the scatterer.

Regarding the actual implementation of these reconfigurable devices, we emphasize how the use of plasma discharges for electromagnetic applications represents a novel and still largely unexplored scientific and engineering area. Plasma, one of the fundamental states of the matter, is known to exhibit a Drude-like electromagnetic behavior, whose plasma frequency can be dynamically modified by acting on the excitation conditions. Plasma discharges confined in elementary shapes, such as cylinders or sphere, have been recently used for implementing different electromagnetic devices, such as antennas and antenna arrays [40]-[44], electromagnetic/photonic band-gap structures [45]-[47] and metamaterials [48]-[51].

The behavior of some of these plasma-based devices is compliant with the technical requirements derived in this work. For instance, in [45] lamps filled with neon at a pressure of about 40 Torr have been used for implementing an electromagnetic band-gap device working in X-band. Considering as a reference the central frequency of this range ($f_0 = 10\ GHz$), we get the following requirements for the Drude-like scatterers proposed in this work: $a = 7.5\ mm$, $f_p^{max} = 37\ GHz$, $\gamma^{max} = 1.2 \times 10^{10} s^{-1}$. These values are in-line with the ones discussed in [45] (and also with the loss factor achieved in [43]), making plasma discharges a suitable candidate for implementing the dipolar-quadrupolar refractive metasurfaces discussed in this work.

## VI. ACKNOWLEDGEMENT

This work has been developed in the frame of the activities of the Project PULSE, funded by the European Innovation Council under the EIC Pathfinder Open 2022 program (protocol number 10109931). Project website is: https://www.pulse-pathfinder.eu/ .